\documentclass[letter,onecolumn,aps,showpacs,showkeys,preprintnumbers,amsmath,amssymb,nofootinbib,superscriptaddress]{revtex4}
\usepackage{epsfig}
\usepackage{amsfonts}
\usepackage{dcolumn}

\newcommand{\newc}{\newcommand}
\newc{\gsim}{\lower.7ex\hbox{$\;\stackrel{\textstyle>}{\sim}\;$}}
\newc{\lsim}{\lower.7ex\hbox{$\;\stackrel{\textstyle<}{\sim}\;$}}
\newc{\gev}{\,{\rm GeV}}
\newc{\mev}{\,{\rm MeV}}
\newc{\ev}{\,{\rm eV}}
\newc{\kev}{\,{\rm keV}}
\newc{\tev}{\,{\rm TeV}}

\newc{\mz}{M_Z}
\newc{\mpl}{M_*}
\newc{\mw}{m_{\rm weak}}
\newc{\nr}[1]{N^c_R{}_{#1}}
\usepackage{amsmath}

\def\beq{\begin{equation}}
\def\eeq{\end{equation}}
\def\bea{\begin{eqnarray}}
\def\eea{\end{eqnarray}}
\def\bi{\begin{itemize}}
\def\ei{\end{itemize}}
\newc{\ie}{{\it i.e.}}          \newc{\etal}{{\it et al.}}
\newc{\eg}{{\it e.g.}}          \newc{\etc}{{\it etc.}}
\newc{\cf}{{\it c.f.}}
\def\bar#1{\overline{#1}}
\def\VEV#1{\left\langle #1 \right\rangle}

\def\inv{^{\raise.15ex\hbox{${\scriptscriptstyle -}$}\kern-.05em 1}}
\def\lbar{{\lower.35ex\hbox{$\mathchar'26$}\mkern-10mu\lambda}} 

\let\<=\langle
\let\>=\rangle

\let\+=\uparrow

\let\ga=\gamma

\let\de=\delta

\let\la=\lambda

\let\si=\sigma

\begin{document}
\author{Anders Basb\o ll}
\email{andersb@phys.au.dk} \affiliation{Department of Physics and
Astronomy, University of Aarhus, Ny Munkegade, DK-8000 Aarhus C}
\title{SUSY Flat Direction Decay - the prospect of particle production and preheating investigated in the unitary gauge}
\date{8th May}


\begin{abstract}
We look at the possibility of non-perturbative particle production
after inflation from SUSY flat directions produced by rotating
eigenstates thereby avoiding the standard adiabaticity conditions.
This might lead to preheating and prevent the delay of
thermalisation of the universe. We investigate the flat directions
$LLE^c$ and $U^cD^cD^c$ and find no particle production. These 2
directions are very important, since they have been named as
possible candidates for being the inflaton. We investigate
$QLQLQLE^c$ and find particle production and therefore the
possibility of preheating. We investigate the $LLE^c$ and
$U^cD^cD^c$ directions appearing simultaneously, and find no
production. Finally, we investigate $LLE^c$ and $QLD^c$
simultaneously - with one L-field in common. Here we do find
particle production and therefore the possibility of preheating.
This means that if SUSY flat directions are to delay thermalisation
and thus explain the (lack of) gravitino production, it is necessary
to explain why complicated directions as $QLQLQLE^c$ are not exited,
and why combinations like $LLE^c$ and $QLD^c$ are not both exited.
\end{abstract}
\maketitle \pagebreak

\section{Introduction}
The scalar potential of the Minimal Supersymmetric Standard Model
(MSSM) possesses a large number of F- and D-flat directions
\cite{Gherghetta:1995dv,Enqvist:2003gh}. These flat directions might
generate the baryon asymmetry of the Universe through the
out-of-equilibrium CP violating decay of coherent field oscillations
along the flat directions themselves
\cite{Affleck:1984fy,Linde:1985gh,Dine:1995uk}.

Recently the cosmological importance of flat direction vacuum
expectation values (VEV)s [Often we will use VEV meaning
\emph{nonzero} vacuum expectation value - this should not cause
confusion] and the decay thereof has been investigated. In
\cite{Allahverdi:2006iq} it was asserted that large flat direction
VEV's can persist long enough to delay thermalization after
inflation and therefore lead to low reheat temperatures. This is of
great importance. A lower reheating temperature would potentially
solve the (lack of) gravitino problem\cite{Allahverdi:2006gralep}.
It has also been claimed \cite{Allahverdi:2006wh} that large flat
direction VEV's can prevent non-perturbative parametric resonant
decay (preheating) of the inflaton since the inflaton decay products
become sufficiently massive preventing preheating from ever becoming
efficient. These arguments hold so long as the flat direction VEV's
do not rapidly decay - they must persist long enough so that they
can delay thermalization and block inflaton preheating. In
\cite{Olive:2006uw} it was claimed that non-perturbative decay can
lead to a rapid depletion of the flat direction condensate and thus
precludes the delay of thermalization after inflation. It was also
concluded that in order for the flat direction to decay
non-perturbatively the system requires more than one flat direction
\cite{Olive:2006uw,Allahverdi:2006xh}. In \cite{Allahverdi:2006xh}
it was stated that even in the presence of multiple flat directions,
some degree of fine-tuning was necessary to achieve flat direction
decay. We note that very recently \cite{Allahverdi:2008} it has been
claimed that even if non-perturbative particle production happen,
the main decay mode will still be perturbative.

The presence of Nambu-Goldstone bosons (Goldstones) is a very
important point in this discussion. The flat directions are charged
under the gauge group of the MSSM. Therefore the flat direction VEV
will break some or all of the gauge symmetries of the theory and
therefore the presence of the associated Goldstones must be
expected. \cite{Olive:2006uw} considers a gauged $U(1)$ model and
constructs the mixing matrix for the excitations around the flat
direction VEV. In \cite{Olive:2006uw} it was claimed that in the
single flat direction case, non-perturbative decay proceeds solely
via a massless Goldstone mode as only the Goldstone mode mixes with
the Higgs and all other massless moduli remain decoupled. Since the
Goldstone represents an unphysical gauge degree of freedom, it was
concluded \cite{Olive:2006uw,Allahverdi:2006xh} that no preheating
occurs in the single flat direction case - a Goldstone can be gauged
away.  In order to determine if flat direction VEV's decay
non-perturbatively into scalar degrees of freedom, one must remove
the Goldstones ie. use the unitary gauge.

In an earlier paper \cite{us} we and our colleagues considered toy
models to demonstrate that, in the unitary gauge, the mixing matrix
of the excitations around a flat direction VEV permits preheating.
Moreover, we found that flat direction decay depends on the number
of dynamical, physical phases appearing in the flat direction VEV.
Specifically, a physical phase difference between two of the
individual field VEV's making up the flat direction is needed.

In the present paper, we look at (some of) the actual SUSY flat
directions. In section \ref{sLLE^c} we look at the $LLE^c$
direction, in section \ref{sNppp} we review the particle production
mechanism from rotating eigenvectors, in section \ref{sLLEcon} we
conclude on the $LLE^c$ case, and in section \ref{sU^cD^cD^c} we do
the $U^cD^cD^c$. The mentioned directions are especially
interesting, since they are mentioned as especially well suited
inflaton candidates in \cite{Allahverdi:2006iq}. We investigate
$QLQLQLE^c$ in section \ref{QLQLQLE^c}. We conclude on one flat
direction in section \ref{sOFDs}. Then we proceed to 2 directions,
first the non-overlapping $LLE^c+U^cD^cD^c$ of the 2 inflaton
candidates of \cite{Allahverdi:2006iq} in section \ref{sUDDLLE} and
then the overlapping directions $QLD^c+LLE^c$ in section
\ref{sQLDLLE}. Finally we look at the simpler approach of just
counting the fields without any calculations in section \ref{scfa}
and conclude in section \ref{scon}.

\section{$LLE^c$}\label{sLLE^c}
One flat direction often mentioned in the literature is $LLE^c$.
Flatness demands the fields with VEV's to come from different
generations, and the 2 L fields with VEV to have opposite
SU(2)-charge. Also, the 3 VEV's must have the same absolute value.
This leaves essentially only 1 choice (when masses are ignored).

We give these VEV's:

\begin{eqnarray}
<\nu_e>=\varphi e^{i\si _1}\nonumber\\
<\mu>=\varphi e^{i\si _2}\\
<\tau^c>=\varphi e^{i\si _3}.\nonumber
\end{eqnarray}
Also, it is clear that 2 other fields will play a role.
\begin{eqnarray}
<e>=0\nonumber\\
<\nu_\mu>=0.
\end{eqnarray}
The Lagrangian reads
\begin{equation}\label{LagrangianLLE^c}
\mathcal{L}=\sum_{i=1}^3\frac{1}{2}|D_{\mu}\Phi_i|^2-V-\frac{1}{4}F_{\mu\nu}^2-\sum_i
\frac{1}{4}W_{\mu\nu}^{i2}
\end{equation}
where for field $\phi_i$
$D_i^{\mu}=\left[(\partial^{\mu}-iq_iA_0^{\mu})\de_{ij}-\sum_{a=1}^3
iP_{ij}^a A_a^\mu\right]\phi_j$ denotes the covariant derivative.
 $P^a$ is the $a^{th}$ Pauli-matrix. The potential we consider arises
from the supersymmetric D-terms and has the form
 \beq
V=\frac{1}{2}\left(D_{H}^2+\sum_a D_a^2\right) \eeq where \bea
D_{H}&=&\frac{g_1}{2}\sum_i q_i |\phi_i|^2 \\
D_a &=&\frac{g_2}{2}\phi^\dag P^a \phi \eea where $q_i$ is the
hypercharge, and $g_1,g_2$ are the hypercharge and SU2 gauge
couplings.

The essential part is removing the Goldstones correctly. To do that
we start by looking at the fields with the VEV's only (no
excitations). We've written those earlier, and we get mixed kinetic
terms
\begin{equation}\label{LkinExpansion}
\mathcal{L}\supset
-\varphi^2A_0(\dot{\si_1}+\dot{\si_2}-2\dot{\si_3})-\varphi^2A_3(\dot{\si_1}-\dot{\si_2})
\end{equation}
which has the form of a coupling between the gauge field and the background condensate. Terms of this type will feed into the equations of motion for the gauge field which, in 
turn, will have an effect on the equations of motion for the scalar
excitations. The remaining terms in $\mathcal{L}$ are \beq
\frac{1}{2}\varphi^2\left[6A_0^2+2A_1^2+2A_2^2+2A_3^2+\dot{\si_1}^2+\dot{\si_2}^2+\dot{\si_3}^2\right]
\eeq - all desired terms.
 By making a $U(1)$ gauge transformation on the VEV,

\beq
\VEV{\Phi_i}\rightarrow\VEV{\Phi^{\prime}_i}=e^{iq_i\la}\VEV{\Phi_i}
\eeq with \beq \la=\frac{2\si_3-\si_1-\si_2}{3}, \eeq and by making
a $SU(2)$ gauge transformation on the VEV,

\beq
\VEV{\Phi_i}\rightarrow\VEV{\Phi^{\prime}_i}=e^{iP^3\ga}\VEV{\Phi_i}
\eeq with \beq \ga=\frac{\si_2-\si_1}{2}, \eeq we can gauge the
unwanted terms away and avoid a complicated analysis of the kinetic
terms. The resulting form of the VEV reads, \bea\label{mod1VEV}
\VEV{ \nu_e}&=&\varphi e^{i\si} \nonumber\\
 \VEV{\mu}&=&\varphi e^{i\si}\\
 \VEV{\tau^c}&=&\varphi e^{i\si}\nonumber
\eea where  $\si=(\si_1+\si_2+\si_3)/3$ represents the remaining
independent physical phase. Following \cite{Kibble}, we can write
the fields in the unitary gauge as (including the other relevant
fields),
 \bea
 \nonumber
 \label{fields21}
 \nu_e&=&(\varphi+\xi_2)e^{i(\si+\frac{\xi_1}{\sqrt{3}\varphi})} \nonumber\\
 e&=&(\xi_5+i\xi_6)e^{i\si} \nonumber\\
\nu_\mu&=&(\xi_7+i\xi_8)e^{i\si} \\
 \mu&=&(\varphi+\xi_3)e^{i(\si+\frac{\xi_1}{\sqrt{3}\varphi})}\nonumber\\
 \tau^c&=&(\varphi+\xi_4)e^{i(\si+\frac{\xi_1}{\sqrt{3}\varphi})}\nonumber
 \eea
where $\sigma$ represents time dependent phase of the VEV (we have
just showed the phase differences are gauged away), $\xi_1$
parameterises its excitation, $\xi_{2,3,4}$ parameterise the
excitations around the VEV, and $\xi_{5,6,7,8}$ parameterise the 2
no-VEV fields (the phase on these fields is not necessary, but
allowed, and will be convenient).

Again we will look at the kinetic term. First, the $\varphi ^2$-term
\beq
\frac{1}{2}\varphi^2\left[6A_0^2+2A_1^2+2A_2^2+2A_3^2+3\dot{\si}^2\right]
\eeq - not surprisingly. This contains no goldstones, so we proceed
to next order.

The terms indicating Goldstones should include $\dot{\xi}_i$. These
terms are
\begin{equation}
\mathcal{L}\supset
-\varphi\left(A_1(\dot{\xi}_6+\dot{\xi}_8)+A_2(\dot{\xi}_7-\dot{\xi}_5)\right).
\end{equation}
The remaining terms are on the forms
\begin{equation}
\mathcal{L}\supset \varphi\left(C_{ijk}\xi_i A_j A_k +D_{ij} \xi_i
A_j \dot{\si}+E_{i}\xi_i \dot{\si}^2-\sqrt{3}\dot{\xi}_1.
\dot{\si}\right)
\end{equation}
Here it is clear, that the field excitation terms (excluding
derivative terms) are suppressed compared to $\varphi ^2$-terms in
the potential. The only excitation is the last term. However, it is
just a "mixing" between an excitation and its own VEV.

The Goldstones are removed by demanding $\dot{\xi}_6=-\dot{\xi}_8$
and $\dot{\xi}_7=\dot{\xi}_5$. Doing this, and renormalising, we
take
 \bea\label{LLE^cDR}
 \nonumber
 \nu_e&=&(\varphi+\xi_2)e^{i(\si+\frac{\xi_1}{\sqrt{3}\varphi})} \nonumber\\
 e&=&\frac{(\xi_5+i\xi_6)}{\sqrt{2}}e^{i\si} \nonumber\\
\nu_\mu&=&\frac{(\xi_5-i\xi_6)}{\sqrt{2}}e^{i\si} \\
 \mu&=&(\varphi+\xi_3)e^{i(\si+\frac{\xi_1}{\sqrt{3}\varphi})}\nonumber\\
 \tau^c&=&(\varphi+\xi_4)e^{i(\si+\frac{\xi_1}{\sqrt{3}\varphi})}.\nonumber
 \eea
This does indeed kill the mixed derivative terms. The remaining
terms stay as they are. But they are all VEV-suppressed, so it is
justified to move to the coordinate derivative, rather than the
covariant derivative.

The remaining kinetic term (to zero'th order in $\varphi$) are
\begin{eqnarray}\label{KinLLE^c}
\nonumber \mathcal{L}\supset \sum_{i=1}^6\left(
\frac{1}{2}\dot{\xi_i}^2\right)+ \sum_{i=2}^6\left(
\frac{1}{2}\xi_i^2\dot{\si}^2\right)\\
+\sum_{i=2}^4 \left(\varphi \xi_i\dot{\si}^2+\frac{2}{\sqrt{3}}\xi_i
\dot{\xi}_1\right)+\frac{3}{2}\varphi ^2\dot{\si}^2+\sqrt{3}\varphi
\dot{\si}\dot{\xi_1}.
\end{eqnarray}
The first term are the kinetic terms that show we have correctly
normalised kinetic fields - including that there are no cross terms.
The second term is completely negligible compared to the $\varphi
^2$-terms of V. The third term, though bigger than the prior one, is
still suppressed. The fourth term is a rotation between the
excitation states. These are very important and will give the
U-matrix below. The fifth term is just a VEV-term, and the final
term is the "mixing" between an excitation and its own VEV. So
everything is fine.

On substituting the fields of eq.\ref{LLE^cDR} into the Lagrangian
given in eq.\ref{LagrangianLLE^c} and defining the vector
$\Xi\equiv(\xi_1,\xi_2,\xi_3,\xi_4,\xi_5,\xi_6)^T$, we find the
quadratic terms
\begin{equation}\label{LagrangianExpanded}
\mathcal{L}\supset\frac{1}{2}|\partial_{\mu}\Xi|^2-\frac{1}{2}\Xi^T
\mathcal{M}^2\Xi-\dot{\Xi}^TU\Xi+...
\end{equation}
where the ellipses denote higher order terms and interactions. The
matrix $U$ given in the second part of the third term in
eq.\ref{KinLLE^c} reads \beq
U_{init}= \left(\begin{array}{cccccc} 0 & -\frac{2\dot{\si}}{\sqrt{3}}& -\frac{2\dot{\si}}{\sqrt{3}}& -\frac{2\dot{\si}}{\sqrt{3}} &0&0\\
0&0&0&0&0&0\\0
&0&0 &0&0&0\\
0&0&0&0&0&0\\0&0&0&0&0&0\\
0&0&0&0&0&0.
\end{array} \right)
\end{equation}
However, we want an antisymmetric matrix for the procedure below.
Using partial integration - and ignoring surface terms - we find
\begin{equation}
U= \left(\begin{array}{cccccc} 0 & -\frac{\dot{\si}}{\sqrt{3}}& -\frac{\dot{\si}}{\sqrt{3}}& -\frac{\dot{\si}}{\sqrt{3}} &0&0\\
\frac{\dot{\si}}{\sqrt{3}}&0&0&0&0&0\\\frac{\dot{\si}}{\sqrt{3}}
&0&0 &0&0&0\\
\frac{\dot{\si}}{\sqrt{3}}&0&0&0&0&0\\0&0&0&0&0&0\\
0&0&0&0&0&0
\end{array} \right)
\end{equation}

while the mass matrix for the physical excitations appears as
\beq\label{MU(1)} \mathcal{M}^2=\varphi^2\left (
\begin{array}{cccccc} 0 & 0 & 0&0&0&0\\ 0& g_1^2+g_2^2  & g_1^2-g_2^2&-2g_1^2&0&0 \\ 0& g_1^2-g_2^2  & g_1^2+g_2^2&-2g_1^2&0&0 \\
0&-2g_1^2&-2g_1^2&4g_1^2& 0  & 0\\0&0&0&0&2g_2^2&0\\0&0&0&0&0&2g_2^2
\end{array} \right)=B\mathcal{M}_d^2B^T,
\eeq
with eigenvalues $M^2_1=6g_1^2\varphi^2$, $M^2_2=M^2_3=M^2_4=2g_2^2\varphi^2$,$M^2_5=M^2_6=0$ (the entries of the diagonal matrix $\mathcal{M}_d$). $B$ is an orthogonal matrix which 
diagonalises $\mathcal{M}^2$ and $M_{1-4}^2$ corresponds to the mass
of the physical eigenstates associated with the spontaneous breaking
of the symmetries. $M^2_5=M^2_6=0$ correspond to the massless
excitations around the flat direction VEV.

The last term in eq.\ref{LagrangianExpanded} appears as a consequence of the
 time-dependence of the background -- it represents a mixing between the fields
  $\xi_{1,2,3,4,5,6}$, 
 and their time-derivatives. The effect of these terms on the system becomes clear if we
 make field redefinitions that remove the mixed derivative terms. 
The resulting transformation leaves the system in an inertial frame
in field space and leads to a time-dependent mass matrix. Defining
$\Xi^\prime=A\Xi$ ($A$ is orthogonal), we find the condition that
$A$ must satisfy in order for all the mixed derivative terms
 to cancel

\begin{equation}
\dot{A}^TA=U.
\end{equation}
The Lagrangian for the $\Xi^\prime$ system now reads
\begin{equation}
\mathcal{L}\supset \frac{1}{2}|\partial_{\mu}\Xi^{\prime}|^2-\frac{1}{2}\Xi^{\prime T} \mathcal{M}^{\prime 2}\Xi^{\prime}
\end{equation}
where $\mathcal{M}^{\prime 2}=A\mathcal{M}^2A^T=AB\mathcal{M}_d^2B^TA^T=C\mathcal{M}_d^{2}C^T$,
and $C=AB$. The matrix $C$ is an orthogonal time-dependent matrix, with columns corresponding to the eigenvectors of $\mathcal{M}^{\prime 2}$. We now have a system of 
scalar fields with canonically normalized kinetic terms and time dependent eigenvectors.

The central point of this discussion centers precisely on the appearance of the {\it time dependent} eigenvectors for the six scalar fields. This satisfies a necessary but not 
sufficient condition for preheating. In the next section, we briefly run through the details of the non-perturbative production of the light scalar fields following the analysis of 
\cite{Nilles:2001fg}. This is a more brief summary than in \cite{us}
- which is otherwise followed here.


\section{Non-perturbative production of particles}\label{sNppp}


Including gravity, the dynamics of the re-scaled conformally coupled scalar fields, $\chi_i=a\Xi^{\prime}_i$,
 where $a$ denotes the scale factor and $\Xi^{\prime}_i$ the $i$-th 
component of the vector $\Xi^{\prime}$, are governed by the
following equations of motion (sum over repeated indices is
implied),
\begin{equation}\label{eommany}
\ddot{\chi_i}+\Omega^2_{ij}(t)\chi_j=0
\end{equation}
where dots represent derivatives with respect to conformal time $t$,
and
\begin{equation}
\label{Omega} \Omega^2_{ij}=a^2
\mathcal{M^{\prime}}_{ij}^2+k^2\delta_{ij},
\end{equation}
where $k$ labels the comoving momentum. Using an orthogonal
time-dependent matrix $C(t)$, we can diagonalise $\Omega_{ij}$ via
$C^T(t)\Omega^2(t)C(t)=\omega^2(t)$, giving the diagonal entries $\omega^2_j(t)$.
 Terms of the form $\sim {\varphi}\dot{\sigma}\dot{\chi}$ arising from the kinetic terms do 
not affect the evolution of the nonzero $k$ quantum modes
\cite{Casadio:2007ip}.

As the vacuum changes, a new set of creation/annihilation operators are required. We use 
Bogolyubov transformation with Bogolyubov coefficients $\alpha$ and
$\beta$ (which denote matrices in the multi-field case).

Initially $\alpha=\mathbb{I}$ and $\beta=0$ while the coupled
differential equations (matrix multiplication implied):
\bea\label{alphadot}
\dot{\alpha} &=& -i \omega\alpha + \frac{\dot{\omega}}{2\omega} \beta - I \alpha - J\beta \nonumber \\
\dot{\beta} &=& \frac{\dot{\omega}}{2\omega} \alpha + i\omega\beta -
J \alpha - I\beta, \label{alandbe} \eea govern the system's time
evolution with the matrices \textit{I} and \textit{J} given by
\begin{equation}
I=\frac{1}{2}\left(\sqrt{\omega}\,C^T\dot{C}\frac{1}{\sqrt{\omega}}+\frac{1}{\sqrt{\omega}}\,C^T\dot{C}\sqrt{\omega}\right)
\end{equation}
\begin{equation}\label{jmatrix}
J=\frac{1}{2}\left(\sqrt{\omega}\,C^T\dot{C}\frac{1}{\sqrt{\omega}}-\frac{1}{\sqrt{\omega}}\,C^T\dot{C}\sqrt{\omega}\right).
\end{equation}
Similarly to the single-field case it can be shown
\cite{Nilles:2001fg} that at any generic time the occupation number
of the $i$th bosonic eigenstate reads (no summation implied)
\begin{equation}\label{n}
n_i(t)=(\beta^*\beta^T)_{ii}.
\end{equation}
As pointed out in \cite{Nilles:2001fg,Olive:2006uw}, there exists two sources of non-adiabaticity in the multi-field scenario. The first source arises from the individual frequency 
time dependence and appears as the only source of non-adiabaticity in the single field case. The second source appears from the time dependence of the frequency matrix 
$\Omega_{ij}$ giving rise to terms in eq.\ref{alandbe} proportional to $I$ and $J$. This second source provides the most important contribution in our analysis and gives rise to 
non-perturbative particle production.

Since initially $\alpha=\mathbb{I}$ and $\beta=0$, eq.\ref{alphadot} shows that a non-vanishing matrix $J$ is a necessary condition to obtain $\dot{\beta}\neq 0$ and hence 
$n_i(t)\neq0$. In general, we have
\begin{equation}\label{Gammaeq}
C^T\dot{C}=B^TA^T\dot{A}B=-B^TUB
\end{equation}
where $A$, $B$ and $U$ were defined in the previous section. The
last equation only holds if \textit{B} is constant in time. This is
obviously the case in the $LLE^c$-case, since, \textit{B}
diagonalises a constant matrix.
\section{$LLE^c$ Conclusion} \label{sLLEcon}
 For the $LLE^c$ example outlined above, $J$ is a $6\times6$ zero matrix.
Therefore there is no particle production and no preheating.

\section{$U^cD^cD^c$}\label{sU^cD^cD^c}
One would expect the $U^cD^cD^c$ case to be much the same -  as
indeed we shall see it is. We give VEV's to these fields (again from
different generations to avoid F-terms)
\begin{eqnarray}
<u^{c\bar{1}}>=\varphi e^{i\si _1}\nonumber\\
<s^{c\bar{2}}>=\varphi e^{i\si _2}\\
<b^{c\bar{3}}>=\varphi e^{i\si _3}.\nonumber
\end{eqnarray}

The Lagrangian reads
\begin{equation}\label{LagrangianU^cD^cD^c}
\mathcal{L}=\sum_{i=1}^3\frac{1}{2}|D_{\mu}\Phi_i|^2-V-\sum_i
\frac{1}{4}F_{\mu\nu}^{i2}-\sum_i \frac{1}{4}G_{\mu\nu}^{i2}
\end{equation}
where for field $\phi_i$
$D_i^{\mu}=\left[(\partial^{\mu}-iq_iA_0^{\mu})\de_{ij}-\sum_{A=1}^8
i GM_{ij}^A B_A^\mu\right]\phi_j$ denotes the covariant derivative.
where $GM^A$ is the $A^{th}$ Gell-Mann-matrix.  The potential now
looks like
 \beq
V=\frac{1}{2}\left(D_{H}^2+\sum_A D_A^2\right) \eeq where \beq D_A
=\frac{g_3}{2}\phi^\dag GM^A \phi \eeq where  $g_3$ is SU(3) gauge
couplings. Removing mixed kinetic terms as before (in 2 tempi), we
use \bea
 \nonumber
 \label{fields21UDD^c2}
 u^{c\bar{1}}&=&(\varphi+\xi_2)e^{i(\si+\frac{\xi_1}{\sqrt{3}\varphi})}, \nonumber\\
 u^{c\bar{2}}&=&\frac{(\xi_5+i\xi_6)}{\sqrt{2}}e^{i\si}, \nonumber\\
 u^{c\bar{3}}&=&\frac{(\xi_7+i\xi_8)}{\sqrt{2}}e^{i\si}, \nonumber\\
 s^{c\bar{1}}&=&\frac{(\xi_5-i\xi_{6})}{\sqrt{2}}e^{i\si}, \nonumber\\
 s^{c\bar{2}}&=&(\varphi+\xi_3)e^{i(\si+\frac{\xi_1}{\sqrt{3}\varphi})}\\
 s^{c\bar{3}}&=&\frac{(\xi_{9}+i\xi_{10})}{\sqrt{2}}e^{i\si}, \nonumber\\
 b^{c\bar{1}}&=&\frac{(\xi_{7}-i\xi_{8})}{\sqrt{2}}e^{i\si}, \nonumber\\
 b^{c\bar{2}}&=&\frac{(\xi_{9}-i\xi_{10})}{\sqrt{2}}e^{i\si}, \nonumber\\
 b^{c\bar{3}}&=&(\varphi+\xi_4)e^{i(\si+\frac{\xi_1}{\sqrt{3}\varphi})}.\nonumber
 \eea

 The U-matrix is

\begin{equation}
U= \left(\begin{array}{cccccccccc} 0 & -\frac{\dot{\si}}{\sqrt{3}}& -\frac{\dot{\si}}{\sqrt{3}}& -\frac{\dot{\si}}{\sqrt{3}} &0&0&0&0&0&0\\
\frac{\dot{\si}}{\sqrt{3}}&0&0&0&0&0&0&0&0&0\\\frac{\dot{\si}}{\sqrt{3}}
&0&0 &0&0&0&0&0&0&0\\
\frac{\dot{\si}}{\sqrt{3}}&0&0&0&0&0&0&0&0&0\\0&0&0&0&0&0&0&0&0&0\\0&0&0&0&0&0&0&0&0&0\\0&0&0&0&0&0&0&0&0&0\\0&0&0&0&0&0&0&0&0&0\\0&0&0&0&0&0&0&0&0&0\\
0&0&0&0&0&0&0&0&0&0
\end{array} \right)
\end{equation}

while the mass matrix for the physical excitations appears as
\beq\label{MU(1)U^cD^cD^c} \mathcal{M}^2=\varphi^2\left (
\begin{array}{cccccccccc} 0 & 0 & 0&0&0&0& 0&0&0&0\\ 0& \frac{16}{9}g_1^2+\frac{4}{3}g_3^2  & \frac{-8}{9}g_1^2+\frac{-2}{3}g_3^2&-\frac{-8}{9}g_1^2+\frac{-2}{3}g_3^2&0&0 \\
 0& \frac{-8}{9}g_1^2+\frac{-2}{3}g_3^2   & \frac{4}{9}g_1^2+\frac{4}{3}g_3^2 &\frac{4}{9}g_1^2+\frac{-2}{3}g_3^2 &0&0&0&0&0&0 \\
0& \frac{-8}{9}g_1^2+\frac{-2}{3}g_3^2   &
\frac{4}{9}g_1^2+\frac{-2}{3}g_3^2
&\frac{4}{9}g_1^2+\frac{4}{3}g_3^2
&0&0&0&0&0&0\\
0&0&0&0&2g_3^2&0&0&0&0&0\\
0&0&0&0&0&2g_3^2&0&0&0&0\\0&0&0&0&0&0&2g_3^2&0&0&0\\0&0&0&0&0&0&0&2g_3^2&0&0\\0&0&0&0&0&0&0&0&2g_3^2&0\\0&0&0&0&0&0&0&0&0&2g_3^2
\end{array} \right)=B\mathcal{M}_d^2B^T,
\eeq with eigenvalues
$M^2_1=\left(\frac{8}{3}g_1^2+2g_3^2\right)\varphi^2$,
$M^2_2=M^2_3=M^2_4=M^2_5=M^2_6=M^2_7=M^2_8=2g_3^2\varphi^2$,$M^2_9=M^2_{10}=0$.
Also here, we end up with $J=0$ and no particle production and
therefore preheating.

\section{$QLQLQLE^c$}\label{QLQLQLE^c}
$QLQLQLE^c$ case has so many fields with nonzero VEV, that all the
phase differences cannot be gauged away. The starting point could be
(notice that here, there are 2 essentially different possibilities -
the squarks having identical $SU(2)$-charge - or not)
\begin{eqnarray}
<u^{c1}>=\varphi e^{i\si _1}\nonumber\\
<c^{c2}>=\varphi e^{i\si _2}\nonumber\\
<t^{c3}>=\varphi e^{i\si _3}\nonumber\\
<e>=\varphi e^{i\si _4}\\
<\mu>=\varphi e^{i\si _5}\nonumber\\
<\tau>=\varphi e^{i\si _6}\nonumber\\
<e^c>=\varphi e^{i\si _7}\nonumber
\end{eqnarray}

The Lagrangian reads
\begin{equation}\label{LagrangianQLQLQLE^c}
\mathcal{L}=\sum_{i=1}^3\frac{1}{2}|D_{\mu}\Phi_i|^2-V-\frac{1}{4}F_{\mu\nu}^2-\frac{1}{4}\sum_i
W_{\mu\nu}^{i2}-\frac{1}{4}\sum_i G_{\mu\nu}^{i2}
\end{equation}
where for field $\phi_i$
$D_i^{\mu}=\left[(\partial^{\mu}-iq_iA_0^{\mu})\de_{ij}-\sum_{a=1}^3
i P_{ij}^a A_a^\mu-\sum_{A=1}^8 i GM_{ij}^A B_A^\mu\right]\phi_j$
denotes the covariant derivative. The potential now looks like
 \beq
V=\frac{1}{2}\left(D_{H}^2+\sum_a D_a^2+\sum_A D_A^2\right). \eeq

 Removing mixed kinetic terms as before (in 2 tempi), we
use \bea
 \nonumber
 \label{fieldsQLQLQLE^c}
u^{c1}&=&(\varphi+\xi_4)e^{i(\si_1+\frac{\xi_1}{\sqrt{7}\varphi})} \nonumber\\
c^{c2}&=&(\varphi+\xi_5)e^{i(\si_1+\frac{\xi_1}{\sqrt{7}\varphi})} \nonumber\\
t^{c3}&=&(\varphi+\xi_6)e^{i(\si_1+\frac{\xi_1}{\sqrt{7}\varphi})} \nonumber\\
e&=&(\varphi+\xi_7)e^{i(\si_1+\frac{\xi_1}{\sqrt{7}\varphi}+\si_2+\frac{\xi_2}{\sqrt{2}\varphi}+\si_3+\frac{\xi_3}{\sqrt{6}\varphi})} \nonumber\\
\mu&=&(\varphi+\xi_8)e^{i(\si_1+\frac{\xi_1}{\sqrt{7}\varphi}-\si_2-\frac{\xi_2}{\sqrt{2}\varphi}+\si_3+\frac{\xi_3}{\sqrt{6}\varphi})} \nonumber\\
\tau&=&(\varphi+\xi_9)e^{i(\si_1+\frac{\xi_1}{\sqrt{7}\varphi}-2\si_3-\frac{2\xi_3}{\sqrt{6}\varphi})} \nonumber\\
e^c&=&(\varphi+\xi_{10})e^{i(\si_1+\frac{\xi_1}{\sqrt{7}\varphi})} \nonumber\\
 u^{c2}&=&\frac{\xi_{11}+i\xi_{12}}{\sqrt{2}}e^{i\si_1} \nonumber\\
 u^{c3}&=&\frac{\xi_{13}+i\xi_{14}}{\sqrt{2}}e^{i\si_1} \nonumber\\
 d^{c1}&=&\left(\frac{\xi_{15}+i\xi_{16}}{\sqrt{2}}
 -\frac{\xi_{21}+i\xi_{22}}{\sqrt{6}}
 -\frac{\xi_{27}+i\xi_{28}}{2\sqrt{3}}
 +\frac{\xi_{29}-i\xi_{30}}{2\sqrt{5}}
 +\frac{\xi_{31}-i\xi_{32}}{\sqrt{30}}\right)e^{i\si_1} \\
c^{c1}&=&\frac{\xi_{11}-i\xi_{12}}{\sqrt{2}}e^{i\si_1} \nonumber\\
c^{c3}&=&\frac{\xi_{19}+i\xi_{20}}{\sqrt{2}}e^{i\si_1} \nonumber\\
s^{c2}&=&\left( \frac{\xi_{21}+i\xi_{22}}{\sqrt{\frac{3}{2}}}
 -\frac{\xi_{27}+i\xi_{28}}{2\sqrt{3}}
 +\frac{\xi_{29}-i\xi_{30}}{2\sqrt{5}}
 +\frac{\xi_{31}-i\xi_{32}}{\sqrt{30}}\right)e^{i\si_1}
 \nonumber\\
t^{c1}&=&\frac{\xi_{13}-i\xi_{14}}{\sqrt{2}}e^{i\si_1} \nonumber\\
t^{c2}&=&\frac{\xi_{19}-i\xi_{20}}{\sqrt{2}}e^{i\si_1} \nonumber\\
b^{c3}&=&\left(\frac{\xi_{27}+i\xi_{28}}{2\sqrt{\frac{1}{3}}}
 +\frac{\xi_{29}-i\xi_{30}}{2\sqrt{5}}
 +\frac{\xi_{31}-i\xi_{32}}{\sqrt{30}}\right)e^{i\si_1}
 \nonumber\\
 \nu_e&=&\left(\frac{\xi_{29}+i\xi_{30}}{2\sqrt{5}}
 -\frac{\xi_{31}+i\xi_{32}}{\sqrt{30}}\right)e^{i(\si_1+\si_2+\si_3)} \nonumber \\
  \nu_{\mu}&=&\left(
 \frac{\xi_{31}+i\xi_{32}}{\sqrt{\frac{6}{5}}}\right)e^{i(\si_1-\si_2+\si_3)}
 \nonumber\\
 \nu_{\tau}&=&\left(\frac{\xi_{15}-i\xi_{16}}{\sqrt{2}}
 +\frac{\xi_{21}-i\xi_{22}}{\sqrt{6}}
 +\frac{\xi_{27}-i\xi_{28}}{2\sqrt{3}}
 -\frac{\xi_{29}+i\xi_{30}}{2\sqrt{5}}
 -\frac{\xi_{31}+i\xi_{32}}{\sqrt{30}}\right)e^{i(\si_1-2\si_3)}.\nonumber
 \eea

 The non-zero elements of the U-matrix are

\begin{eqnarray}
U_{4,1}&=&U_{5,1}=U_{6,1}=U_{10,1}=\frac{\si_1'}{\sqrt{7}}\nonumber\\
U_{7,1}&=&=\frac{\si_1'+\si_2'+\si_3'}{\sqrt{7}}\nonumber\\
U_{7,2}&=&=\frac{\si_1'+\si_2'+\si_3'}{\sqrt{2}}\nonumber\\
U_{7,3}&=&=\frac{\si_1'+\si_2'+\si_3'}{\sqrt{6}}\nonumber\\
U_{8,1}&=&\frac{\si_1'-\si_2'+\si_3'}{\sqrt{7}}\nonumber\\
U_{8,2}&=&\frac{\si_1'-\si_2'+\si_3'}{\sqrt{2}}\nonumber\\
U_{8,3}&=&\frac{\si_1'-\si_2'+\si_3'}{\sqrt{6}}\nonumber\\
U_{9,1}&=&\frac{\si_1'-2\si_3'}{\sqrt{7}}\nonumber\\
U_{3,9}&=&\frac{\si_1'-2\si_3'}{\sqrt{\frac{3}{2}}}\nonumber\\
U_{15,16}&=&\si_3'\nonumber\\
U_{20,15}&=&U_{16,19}=\frac{\si_1'-\si_3'}{\sqrt{3}}\nonumber\\
U_{22,15}&=&U_{16,21}=\frac{\si_1'-\si_3'}{\sqrt{6}}\nonumber\\
U_{24,15}&=&U_{23,16}=\frac{\si_1'-\si_3'}{\sqrt{10}}\\
U_{26,15}&=&U_{25,16}=\frac{\si_1'-\si_3'}{\sqrt{15}}\nonumber\\
U_{19,20}&=&\frac{2\si_1'+\si_3'}{3}\nonumber\\
U_{22,19}&=&U_{20,21}=\frac{\si_1'-\si_3'}{3\sqrt{2}}\nonumber\\
U_{24,19}&=&U_{23,20}=\frac{\si_1'-\si_3'}{\sqrt{30}}\nonumber\\
U_{26,19}&=&U_{25,20}=\frac{\si_1'-\si_3'}{3\sqrt{5}}\nonumber\\
U_{21,22}&=&\frac{5\si_1'+\si_3'}{6}\nonumber\\
U_{24,21}&=&U_{23,22}=\frac{\si_1'-\si_3'}{2\sqrt{15}}\nonumber\\
U_{26,21}&=&U_{25,22}=\frac{\si_1'-\si_3'}{3\sqrt{10}}\nonumber\\
U_{23,24}&=&\frac{7\si_1'+8\si_2'+7\si_3'}{10}\nonumber\\
U_{26,23}&=&U_{24,25}=\frac{3\si_1'+2\si_2'+3\si_3'}{5\sqrt{6}}\nonumber\\
U_{25,26}&=&\frac{4\si_1'-4\si_2'+4\si_3'}{5}\nonumber
\end{eqnarray}
and their antisymmetric counterparts.

The mass matrix for the physical excitations appears as (in units of
$\varphi^2$) \begin{eqnarray}
M_{4,4}&=&M_{5,5}=M_{6,6}=\frac{g_1^2}{9}+g_2^2+\frac{4g_3^2}{3}\nonumber\\
M_{4,5}&=&M_{4,6}=M_{5,6}=\frac{g_1^2}{9}+g_2^2-\frac{2g_3^2}{3}\nonumber\\
M_{4,7}&=&M_{4,8}=M_{4,9}=M_{5,7}=M_{5,8}=M_{5,9}=
M_{6,7}=M_{6,8}=M_{6,9}=\frac{-g_1^2}{3}-g_2^2\nonumber\\
M_{4,10}&=&M_{5,10}=\frac{2g_1^2}{3}\nonumber\\
M_{7,7}&=&M_{7,8}=M_{7,9}=M_{8,8}=M_{8,9}=M_{9,9}=g_1^2+g_2^2\nonumber\\
M_{7,10}&=&=M_{8,10}=M_{9,10}=-2g_1^2\nonumber\\
M_{10,10}&=&4g_1^2\nonumber\\
M_{11,11}&=&M_{12,12}=M_{13,13}=M_{14,14}=M_{17,17}=M_{18,18}=2g_3^2\nonumber\\
M_{15,15}&=&M_{16,16}=2g_2^2\nonumber\\
M_{15,19}&=&M_{16,20}=\frac{2g_2^2}{\sqrt{3}}\nonumber\\
M_{15,21}&=&M_{16,22}=\frac{\sqrt{2}g_2^2}{\sqrt{3}}\nonumber\\
M_{15,23}&=&-M_{16,24}=\frac{3\sqrt{2}g_2^2}{\sqrt{5}}\\
M_{15,25}&=&-M_{16,26}=\frac{2\sqrt{3}g_2^2}{\sqrt{5}}\nonumber\\
M_{19,19}&=&M_{20,20}=\frac{2g_2^2}{3}\nonumber\\
M_{19,21}&=&M_{20,22}=\frac{\sqrt{2}g_2^2}{3}\nonumber\\
M_{19,23}&=&-M_{20,24}=\frac{\sqrt{6}g_2^2}{\sqrt{5}}\nonumber\\
M_{19,25}&=&-M_{20,26}=\frac{2g_2^2}{\sqrt{5}}\nonumber\\
M_{21,21}&=&M_{22,22}=\frac{g_2^2}{3}\nonumber\\
M_{21,23}&=&-M_{22,24}=\frac{\sqrt{3}g_2^2}{\sqrt{5}}\nonumber\\
M_{21,25}&=&-M_{22,26}=\frac{\sqrt{2}g_2^2}{\sqrt{5}}\nonumber\\
M_{23,23}&=&M_{24,24}=\frac{9g_2^2}{5}\nonumber\\
M_{23,25}&=&M_{24,26}=\frac{3\sqrt{6}g_2^2}{5}\nonumber\\
M_{25,25}&=&M_{26,26}=\frac{6g_2^2}{5}\nonumber
\end{eqnarray}
and their symmetric counterparts. The eigenvalues are
$M^2_1=\frac{11g_1^2+9g_2^2+\sqrt{121g_1^4-54g_1^2g_2^2+81g_2^4}}{3}\varphi^2$,
$M^2_2=\frac{11g_1^2+9g_2^2-\sqrt{121g_1^4-54g_1^2g_2^2+81g_2^4}}{3}\varphi^2$,$M^2_3=M^2_4=6g_2^2\varphi^2,
M^2_5=M^2_6=M^2_7=M^2_8=M^2_9=M^2_{10}=M^2_{11}=M^2_{12}=2g_3^2\varphi^2$,$M^2_{13}=....=M^2_{26}=0$.

The \textit{J}-matrix is (really: The J-matrix can be – the
splitting of eigenspaces of higher dimensions into subspaces is
arbitrary)
\begin{eqnarray}
J_{4,13}&=&-J_{3,14}=\frac{-\sqrt{k}+\sqrt{k+\frac{6g_2^2\varphi
^2}{k}}}{4\sqrt{2}(k^2+6g_2^2\varphi ^2)^\frac{1}{4}}
\left(\si_2 '-2\si_3 '\right)\nonumber\\
J_{3,16}&=&-J_{4,15}=\frac{-\sqrt{k}+\sqrt{k+\frac{6g_2^2\varphi
^2}{k}}}{4\sqrt{30}(k^2+6g_2^2\varphi ^2)^\frac{1}{4}}
\left(5\si_2 '+6\si_3 '\right)\nonumber\\
J_{4,17}&=&J_{3,18}=-\frac{\sqrt{3}(-\sqrt{k}+\sqrt{k+\frac{6g_2^2\varphi
^2}{k}})}{4\sqrt{10}(k^2+6g_2^2\varphi ^2)^\frac{1}{4}}
\si_3 '\nonumber\\
J_{4,19}&=&J_{3,20}=-\frac{-\sqrt{k}+\sqrt{k+\frac{6g_2^2\varphi
^2}{k}}}{4\sqrt{2}(k^2+6g_2^2\varphi ^2)^\frac{1}{4}}
\si_3 '\nonumber\\
J_{1,24}&=&\frac{\left(g_1^2+9g_2^2+\sqrt{121g_1^4-54g_1^2g_2^2+81g_2^4}\right)
\left(-3\sqrt{k}+\sqrt{\frac{9k^2+3(11g_1^2+9g_2^2+\sqrt{121g_1^4-54g_1^2g_2^2+81g_2^4})\varphi
^2}{k}}\right)}{3^\frac{3}{4}8\sqrt{14}g_1^2\sqrt{\frac{\sqrt{121g_1^4-54g_1^2g_2^2+81g_2^4}}{3g_1^2-9g_2^2+\sqrt{121g_1^4-54g_1^2g_2^2+81g_2^4}}}
\left(3k^2+\left(11g_1^2+9g_2^2+\sqrt{121g_1^4-54g_1^2g_2^2+81g_2^4}\right)\varphi
^2\right)^\frac{1}{4}}\si_3 '\\
J_{1,25}&=&\frac{\left(g_1^2+9g_2^2+\sqrt{121g_1^4-54g_1^2g_2^2+81g_2^4}\right)
\left(-3\sqrt{k}+\sqrt{\frac{3k^2+3(11g_1^2+9g_2^2+\sqrt{121g_1^4-54g_1^2g_2^2+81g_2^4})\varphi
^2}{k}}\right)}{3^\frac{3}{4}8\sqrt{14}g_1^2\sqrt{\frac{\sqrt{121g_1^4-54g_1^2g_2^2+81g_2^4}}{3g_1^2-9g_2^2+\sqrt{121g_1^4-54g_1^2g_2^2+81g_2^4}}}
\left(3k^2+\left(11g_1^2+9g_2^2+\sqrt{121g_1^4-54g_1^2g_2^2+81g_2^4}\right)\varphi
^2\right)^\frac{1}{4}}\si_2 '\nonumber\\
J_{2,24}&=&\frac{\left(g_1^2+9g_2^2-\sqrt{121g_1^4-54g_1^2g_2^2+81g_2^4}\right)
\left(-3\sqrt{k}+\sqrt{\frac{9k^2+3(11g_1^2+9g_2^2-\sqrt{121g_1^4-54g_1^2g_2^2+81g_2^4})\varphi
^2}{k}}\right)}{3^\frac{3}{4}8\sqrt{14}g_1^2\sqrt{\frac{\sqrt{121g_1^4-54g_1^2g_2^2+81g_2^4}}{-3g_1^2+9g_2^2+\sqrt{121g_1^4-54g_1^2g_2^2+81g_2^4}}}
\left(3k^2+\left(11g_1^2+9g_2^2-\sqrt{121g_1^4-54g_1^2g_2^2+81g_2^4}\right)\varphi
^2\right)^\frac{1}{4}}\si_3 '\nonumber\\
J_{2,25}&=&\frac{\left(g_1^2+9g_2^2-\sqrt{121g_1^4-54g_1^2g_2^2+81g_2^4}\right)
\left(-3\sqrt{k}+\sqrt{\frac{3k^2+3(11g_1^2+9g_2^2-\sqrt{121g_1^4-54g_1^2g_2^2+81g_2^4})\varphi
^2}{k}}\right)}{3^\frac{3}{4}8\sqrt{14}g_1^2\sqrt{\frac{\sqrt{121g_1^4-54g_1^2g_2^2+81g_2^4}}
{-3g_1^2+9g_2^2+\sqrt{121g_1^4-54g_1^2g_2^2+81g_2^4}}}
\left(3k^2+\left(11g_1^2+9g_2^2-\sqrt{121g_1^4-54g_1^2g_2^2+81g_2^4}\right)\varphi
^2\right)^\frac{1}{4}}\si_2 '\nonumber
\end{eqnarray}
and their symmetric counterparts.

Here the J matrix show rotation between states 1-4 and the light
states, giving particle production and possible preheating. The
reason that the SU(3) states do not rotate is that the 3 Q's have
the same SU(2)-charge, and the 2 diagonal SU(3) generators have
removed the phases between them.

In fact, changing the assignments such that the quarks have split
SU(2)-charges will change something, even the eigenvalues, but it
will not change that J is nonzero and preheating is possible.

\section{One Flat direction - summary}\label{sOFDs}

For the 2 flat directions mentioned as the most obvious candidate to
be the inflaton in \cite{Allahverdi:2006iq}, $QLD^c$ and $LLE^c$, we
find no preheating. The reason \cite{us} found differently with a
toy model direction of 3 superfields was that it was rather special
to have 3 VEV-fields and only 1 broken generator. When only 1
generator was broken, only one phase difference was removed, and the
second phase difference gave the preheating. However, for $LLE^c$
and $U^cD^cD^c$ 2 diagonal generators are broken and there is no
preheating due to the diagonal generators. We think, inspired by
\cite{Olive:2006uw}, it makes sense to split the involved fields in
those connected to VEV's by the diagonal generators (from here:
Sector 1), and those connected to the VEV by the off-diagonal
generators (from here: Sector 2). In this case, and we suspect in
most others, the structure of Sector 2, is that the massive states
are Higgses, and they all have the same eigenvalue. Therefore
rotation does not have an effect (in fact, rotation does not make
sense, since one cannot distinguish the eigenstates). In $QLD^c$
though, they have different eigenvalues - some fields connected to
the VEV through $SU(2)$, others through $SU(3)_c$. However, for each
field it is either or. Any difference from this, should be if a
field is connected to 2 VEV's, one by a $SU(2)$ and one by a
$SU(3)_c$ generator. For Sector 1, it would take more than 3 fields
(or less than 2 broken generators). This is what happens in
$QLQLQLE^c$. We can gauge away 4 phase differences, but this leaves
2 phase differences that can give the preheating. There could also
be a mixing between sectors, if a field was connected to 1 VEV by a
diagonal generator and to another by an off-diagonal one. However,
this seems impossible for a single flat direction.
\section{$U^cD^cD^c$, $LLE^c$ simultaniously}\label{sUDDLLE}
The 2 directions first presented can co-exist. In fact, there is no
reason why they should not both get large VEV's \cite{Olive:2006uw}.
It is not so easy to argue why there should be no preheating - since
now we have 6 Sector 1 fields, and only 4 diagonal generators to
break.
\begin{eqnarray}
<u^{c\bar{1}}>=\varphi e^{i\si _1}\nonumber\\
<s^{c\bar{2}}>=\varphi e^{i\si _2}\nonumber\\
<b^{c\bar{3}}>=\varphi e^{i\si _3}\nonumber\\
<\nu_e>=A\varphi e^{i\si _4}\\
<\mu>=A\varphi e^{i\si _5}\nonumber\\
<\tau^c>=A\varphi e^{i\si _6}\nonumber
\end{eqnarray}
where A is the relation between the absolute value of the VEV's
involved. The Lagrangian reads
\begin{equation}\label{LagrangianU^cD^cD^cLLE^c}
\mathcal{L}=\sum_{i=1}^3\frac{1}{2}|D_{\mu}\Phi_i|^2-V-\frac{1}{4}F_{\mu\nu}^2-\frac{1}{4}\sum_i
W_{\mu\nu}^{i2}-\frac{1}{4}\sum_i G_{\mu\nu}^{i2}
\end{equation}
where for field $\phi_i$
$D_i^{\mu}=\left[(\partial^{\mu}-iq_iA_0^{\mu})\de_{ij}-\sum_{a=1}^3
i P_{ij}^a A_a^\mu-\sum_{A=1}^8 i GM_{ij}^A B_A^\mu\right]\phi_j$
denotes the covariant derivative. The potential now looks like
 \beq
V=\frac{1}{2}\left(D_{H}^2+\sum_a D_a^2+\sum_A D_A^2\right). \eeq

To remove mixed kinetic terms
 we must reparametrise \bea
\label{fields21U^cD^cD^cLLE^c4}
u^{c\bar{1}}&=&(\varphi+\xi_2)e^{i(\si_1+\frac{\xi_1}{\sqrt{3}\varphi})},
\nonumber\\
u^{c\bar{2}}&=&\frac{\xi_5+i\xi_6}{\sqrt{2}}e^{i\si_1}, \nonumber\\
u^{c\bar{3}}&=&\frac{\xi_7+i\xi_8}{\sqrt{2}}e^{i\si_1}, \nonumber\\
s^{c\bar{1}}&=&\frac{\xi_5-i\xi_6}{\sqrt{2}}e^{i\si_1}, \nonumber\\
s^{c\bar{2}}&=&(\varphi+\xi_3)e^{i(\si_1+\frac{\xi_1}{\sqrt{3}\varphi})}\nonumber\\
s^{c\bar{3}}&=&\frac{\xi_{9}+i\xi_{10}}{\sqrt{2}}e^{i\si_1}, \nonumber\\
b^{c\bar{1}}&=&\frac{\xi_7-i\xi_8}{\sqrt{2}}e^{i\si_1}, \nonumber\\
b^{c\bar{2}}&=&\frac{\xi_9-i\xi_{10}}{\sqrt{2}}e^{i\si_1}, \\
b^{c\bar{3}}&=&(\varphi+\xi_4)e^{i(\si_1+\frac{\xi_1}{\sqrt{3}\varphi})}\nonumber\\
\nu_e&=&(A\varphi+\xi_{12})e^{i(\si_2+\frac{\xi_{11}}{\sqrt{3}A\varphi})},
\nonumber\\
e&=&\frac{\xi_{15}+i\xi_{16}}{\sqrt{2}}e^{i\si_2}, \nonumber\\
\nu_\mu&=&\frac{\xi_{15}-i\xi_{16}}{\sqrt{2}}e^{i\si_2}, \nonumber\\
\mu&=&(A\varphi+\xi_{13})e^{i(\si_2+\frac{\xi_{11}}{\sqrt{3}A\varphi})}\nonumber\\
\tau^c&=&(A\varphi+\xi_{14})e^{i(\si_2+\frac{\xi_{11}}{\sqrt{3}A\varphi})}\nonumber
\eea After verifying that the mixed derivatives have indeed been
removed, we use coordinate derivatives, and find the remaining
kinetic terms (those not to second order) to be: (to zero'th order
in $\varphi$)
\begin{eqnarray}
\nonumber \mathcal{L}&\supset& \sum_{i=1}^{16}\left(
\frac{1}{2}\dot{\xi_i}^2\right)+ \sum_{i=2}^{10}\left(
\frac{1}{2}\xi_i^2\dot{\si_1}^2\right)+ \sum_{i=12}^{16}\left(
\frac{1}{2}\xi_i^2\dot{\si_2}^2\right)\\
&&+\sum_{i=2}^4 \left(\varphi
\xi_i\dot{\si_1}^2+\frac{2}{\sqrt{3}}\xi_i
\dot{\xi}_1\right)+\sum_{i=12}^{14} \left(A\varphi
\xi_i\dot{\si_2}^2+\frac{2}{\sqrt{3}}\xi_i
\dot{\xi}_{11}\right)\\
&&+\frac{3}{2}\varphi ^2\dot{\si_1}^2+\frac{3}{2}A^2\varphi
^2\dot{\si_2}^2+\sqrt{3}\varphi
\dot{\si_1}\dot{\xi_1}+\sqrt{3}A\varphi
\dot{\si_2}\dot{\xi_{11}}.\nonumber
\end{eqnarray}
It seems the 2 directions do not "see" each other. We find (after
antisymmetrising)
\begin{equation}
U= \left(\begin{array}{cccccccccccccccc} 0 & -\frac{\dot{\si_1}}{\sqrt{3}}& -\frac{\dot{\si_1}}{\sqrt{3}}& -\frac{\dot{\si_1}}{\sqrt{3}} &0&0&0&0&0&0&0&0&0&0&0&0\\
\frac{\dot{\si_1}}{\sqrt{3}}&0&0&0&0&0&0&0&0&0&0&0&0&0&0&0\\\frac{\dot{\si_1}}{\sqrt{3}}
&0&0 &0&0&0&0&0&0&0&0&0&0&0&0&0\\
\frac{\dot{\si_1}}{\sqrt{3}}&0&0&0&0&0&0&0&0&0&0&0&0&0&0&0\\0&0&0&0&0&0&0&0&0&0&0&0&0&0&0&0\\0&0&0&0&0&0&0&0&0&0&0&0&0&0&0&0\\0&0&0&0&0&0&0&0&0&0&0&0&0&0&0&0\\0&0&0&0&0&0&0&0&0&0&0&0&0&0&0&0\\
0&0&0&0&0&0&0&0&0&0&0&0&0&0&0&0\\
0&0&0&0&0&0&0&0&0&0&0&0&0&0&0&0\\0 &0&0&0&0&0&0&0&0&0&0& -\frac{\dot{\si_2}}{\sqrt{3}}& -\frac{\dot{\si_2}}{\sqrt{3}}& -\frac{\dot{\si_2}}{\sqrt{3}} &0&0\\
0&0&0&0&0&0&0&0&0&0&\frac{\dot{\si_2}}{\sqrt{3}}&0&0&0&0&0\\
0&0&0&0&0&0&0&0&0&0&\frac{\dot{\si_2}}{\sqrt{3}}
&0&0 &0&0&0\\
0&0&0&0&0&0&0&0&0&0&\frac{\dot{\si_2}}{\sqrt{3}}&0&0&0&0&0\\0&0&0&0&0&0&0&0&0&0&0&0&0&0&0&0\\0&0&0&0&0&0&0&0&0&0&0&0&0&0&0&0\nonumber
\end{array} \right)
\end{equation}
This looks as if the 2 parts are completely separated. The mass
matrix for the physical excitations appears as
\beq\label{MU(1)U^cD^cD^cLLE^c} \mathcal{M}^2=\varphi^2
\end{equation}
\small{ \beq\left (
\begin{array}{cccccccccccccccc}0&0&0&0&0&0&0&0&0&0&0&0&0&0&0&0\\
 0& \frac{16g_1^2+12g_3^2}{9}  & \frac{-8g_1^2-6g_3^2}{9}&\frac{8g_1^2+6g_3^2}{9}&0&0&0&0&0&0&0& \frac{4Ag_1^2}{3}
&\frac{4Ag_1^2}{3}
&\frac{-8Ag_1^2}{3} &0&0\\
 0& \frac{-8g_1^2-6g_3^2}{9}& \frac{4g_1^2+12g_3^2}{9}&\frac{4g_1^2-6g_3^2}{9}&0&0&0&0&0&0
 &0&\frac{-2Ag_1^2}{3}
&\frac{-2Ag_1^2}{3}
&\frac{4Ag_1^2}{3} &0&0 \\
0& \frac{-8g_1^2-6g_3^2}{9} &\frac{4g_1^2-6g_3^2}{9}
&\frac{4g_1^2+12g_3^2}{9}&0&0&0&0&0&0&0& \frac{-2Ag_1^2}{3}
&\frac{-2Ag_1^2}{3}
&\frac{4Ag_1^2}{3} &0&0\\
0&0&0&0&2g_3^2&0&0&0&0&0&0&0&0&0&0&0\\
0&0&0&0&0&2g_3^2&0&0&0&0&0&0&0&0&0&0\\
0&0&0&0&0&0&2g_3^2&0&0&0&0&0&0&0&0&0\\
0&0&0&0&0&0&0&2g_3^2&0&0&0&0&0&0&0&0\\
0&0&0&0&0&0&0&0&2g_3^2&0&0&0&0&0&0&0\\
0&0&0&0&0&0&0&0&0&2g_3^2&0&0&0&0&0&0\\0&0&0&0&0&0&0&0&0&0&0&0&0&0&0&0\\
0& \frac{4Ag_1^2}{3} &\frac{-2Ag_1^2}{3} &\frac{-2Ag_1^2}{3}
&0&0&0&0&0&0&0&A^2g_1^2+A^2g_2^2&A^2g_1^2-A^2g_2^2&-2A^2g_1^2&0&0\\
0& \frac{4Ag_1^2}{3} &\frac{-2Ag_1^2}{3} &\frac{-2Ag_1^2}{3}
&0&0&0&0&0&0&0&A^2g_1^2-A^2g_2^2&A^2g_1^2+A^2g_2^2&-2A^2g_1^2&0&0\\
0& \frac{-8Ag_1^2}{3} &\frac{4Ag_1^2}{3} &\frac{4Ag_1^2}{3}
&0&0&0&0&0&0&0&-2A^2g_1^2&-2A^2g_1^2&4A^2g_1^2&0&0\\
0&0&0&0&0&0&0&0&0&0&0&0&0&0&2A^2g_2^2&0\\
0&0&0&0&0&0&0&0&0&0&0&0&0&0&0&2A^2g_3^2
\end{array} \right)
\eeq} with eigenvalues \bea
M^2_1&=&\left(\frac{4+9A^2}{3}g_1^2+g_3^2+\frac{1}{3}\sqrt{(4+9A^2)^2g_1^4-6(-4+9A^2)g_1^2g_3^2+9g_3^4}\right)\varphi^2\nonumber\\
 M^2_2&=&\left(\frac{4+9A^2}{3}g_1^2+g_3^2-\frac{1}{3}\sqrt{(4+9A^2)^2g_1^4-6(-4+9A^2)g_1^2g_3^2+9g_3^4}\right)\varphi^2\nonumber\\
 M^2_3&=&M^2_4=M^2_5=M^2_6=M^2_7=M^2_8=M^2_9=2g_3^2\varphi^2\\
 M^2_{10}&=&M^2_{11}=M^2_{12}=2A^2g_2^2\varphi^2\nonumber\\
 M^2_{13}&=&M^2_{14}=M^2_{15}=M^2_{16}=0.\nonumber\eea

Even though this indeed looks like a mixing between the two
directions, again $J=0$ (16 by 16 matrix) and there is no particle
production and therefore no preheating.

\section{$QLD^c+LLE^c$ - an overlapping direction}\label{sQLDLLE}
While $LLE^c+QLD^c$ shows that more than 1 phase is not enough to
secure preheating, it is clear that the 2 directions did not
overlap. There are flat directions more intimately connected. One
example of this is $QLD^c+LLE^c$ - with one $L$-field in common.
This is exiting, since here the flat directions cannot just have one
phase each. $Q$ and $D^c$ must be from different generations and
have the same (or opposite, if you like) colour charge. It is easy
to show that flatness is independent of phase, and that the common
field shall have a VEV that is the square root of the sum of squares
of the VEV's from the 2 directions.

\begin{eqnarray}
<d^{c^{1}}>=A\varphi e^{i\si _4}\nonumber\\
<s^{c\bar{1}}>=A\varphi e^{i\si _5}\nonumber\\
<\nu_e>=\sqrt{1+A^2}\varphi e^{i\si _3}\\
<\mu>=\varphi e^{i\si _2}\nonumber\\
<\tau^c>=\varphi e^{i\si _1}\nonumber
\end{eqnarray}
where A is the relation between the absolute values of the VEV's.
The Lagrangian, covariant derivatives and the potential looks as
before.

To remove mixed kinetic terms, we must re-parameterise
 \bea
\nonumber \label{fields21QLDLLE^c2}
u^{c^{1}}&=&\left(\frac{\xi_8+i\xi_9}{\sqrt{1+A^2}}+\frac{A(\xi_{14}-i\xi_{15})}{\sqrt{2(1+A^2)}}\right)
e^{i\left(\si+\left(-1-\frac{1-A^2}{5(1+A^2)}\right)\ga\right)} \nonumber\\
d^{c^{1}}&=&(A\varphi+\xi_4)
e^{i\left((\si+\frac{1}{\sqrt{5}}\xi€_6)+\left(-1-\frac{1-A^2}{5(1+A^2)}\right)\left(\ga+\frac{\sqrt{5}(1+A^2)}{2\sqrt{6+8A^2+6A^4}}\xi_7\right)\right)}\nonumber\\
d^{c^{2}}&=&\frac{\xi_{10}+i\xi_{11}}{\sqrt{2}}e^{i\left(\si+\left(-1-\frac{1-A^2}{5(1+A^2)}\right)\ga\right)}
\nonumber\\
d^{c^{3}}&=&\frac{\xi_{12}+i\xi_{13}}{\sqrt{2}}e^{i\left(\si+\left(-1-\frac{1-A^2}{5(1+A^2)}\right)\ga\right)}
\nonumber\\
s^{c\bar{1}}&=&(A\varphi+\xi_5)
e^{i\left((\si+\frac{1}{\sqrt{5}}\xi_6)+\left(-1-\frac{1-A^2}{5(1+A^2)}\right)\left(\ga+\frac{\sqrt{5}(1+A^2)}{2\sqrt{6+8A^2+6A^4}}\xi_7\right)\right)}\nonumber\\
s^{c\bar{2}}&=&\frac{-\xi_{10}+i\xi_{11}}{\sqrt{2}}e^{i\left(\si+\left(-1-\frac{1-A^2}{5(1+A^2)}\right)\ga\right)}
\\
s^{c\bar{3}}&=&\frac{-\xi_{12}+i\xi_{13}}{\sqrt{2}}e^{i\left(\si+\left(-1-\frac{1-A^2}{5(1+A^2)}\right)\ga\right)}
\nonumber\\
\nu_e&=&(\sqrt{1+A^2}\varphi+\xi_3)
e^{i\left((\si+\frac{1}{\sqrt{5}}\xi_6)+\frac{4(1-A^2)}{5(1+A^2)}\left(\ga+\frac{\sqrt{5}(1+A^2)}{2\sqrt{6+8A^2+6A^4}}\xi_7\right)\right)}\nonumber\\
e&=&\frac{\xi_{14}+i\xi_{15}}{\sqrt{2}}e^{i\left(\si+\frac{4(1-A^2)}{5(1+A^2)}\ga\right)}
\nonumber\\
\nu_\mu&=&\left(\frac{-A(\xi_{8}+i\xi_{9})}{\sqrt{1+A^2}}+\frac{\xi_{14}-i\xi_{15}}{\sqrt{2(1+A^2)}}\right)
e^{i\left(\si+\left(1-\frac{1-A^2}{5(1+A^2)}\right)\ga\right)} \nonumber\\
\mu&=&(\varphi+\xi_2)
e^{i\left((\si+\frac{1}{\sqrt{5}}\xi_6)+\left(1-\frac{1-A^2}{5(1+A^2)}\right)\left(\ga+\frac{\sqrt{5}(1+A^2)}{2\sqrt{6+8A^2+6A^4}}\xi_7\right)\right)}\nonumber\\
\nonumber \tau^c&=&(\varphi+\xi_1)
e^{i\left((\si+\frac{1}{\sqrt{5}}\xi_6)+\left(1-\frac{1-A^2}{5(1+A^2)}\right)\left(\ga+\frac{\sqrt{5}(1+A^2)}{2\sqrt{6+8A^2+6A^4}}\xi_7\right)\right)}
\eea

There are quite many kinetic terms now, but they include
\begin{eqnarray}
\mathcal{L}&\supset& \sum_{i=1}^{15}\left(
\frac{1}{2}\dot{\xi_i}^2\right)
\end{eqnarray}
and no cross terms. The U-matrix is (after antisymmetrising) \bea
U_{1,6}&=&-U_{6,1}=U_{2,6}=-U_{6,2}=\frac{(4+6A^2)\dot{\ga}+5(1+A^2)\dot{\si}}{5\sqrt{5}(1+A^2)}\nonumber\\
U_{1,7}&=&-U_{7,1}=U_{2,7}=-U_{7,2}=\frac{(2+3A^2)((4+6A^2)\dot{\ga}+5(1+A^2)\dot{\si})}{5\sqrt{10}(1+A^2)\sqrt{3+4A^2+3A^4}}\nonumber\\
U_{3,6}&=&-U_{6,3}=\frac{4(1-A^2)\dot{\ga}+5(1+A^2)\dot{\si}}{5\sqrt{5}(1+A^2)}\nonumber\\
U_{3,7}&=&-U_{7,3}=\frac{\sqrt{2}(-1+A^2)(4(-1+A^2)\dot{\ga}-5(1+A^2)\dot{\si})}{5\sqrt{5}(1+A^2)\sqrt{3+4A^2+3A^4}}\nonumber\\
U_{4,6}&=&-U_{6,4}=U_{5,6}=-U_{6,5}=\frac{-(6+4A^2)\dot{\ga}+5(1+A^2)\dot{\si}}{5\sqrt{5}(1+A^2)}\\
U_{4,7}&=&-U_{7,4}=U_{5,7}=-U_{7,5}=\frac{(3+2A^2)((6+4A^2)\dot{\ga}-5(1+A^2)\dot{\si})}{5\sqrt{10}(1+A^2)\sqrt{3+4A^2+3A^4}}\nonumber\\
U_{8,9}&=&-U_{9,8}=\frac{6(-1+A^2)\dot{\ga}}{5(1+A^2)}+\dot{\si}\nonumber\\
U_{8,15}&=&-U_{15,8}=U_{9,14}=-U_{14,9}=\frac{\sqrt{2}A\dot{\ga}}{(1+A^2)}\nonumber\\
U_{remaining}&=&0.\nonumber\eea
 This
looks as if everything mixes - certainly it does not look as if
there are 2 separable parts. The mass matrix is quite complicated,
but the structure is like this \beq\label{MU(1)QLDLLE^c}
\mathcal{M}^2=\varphi^2\left (
\begin{array}{cc} M_{7\times 7}&0_{7\times 8}\\
 0_{8\times 7}&D_{8\times 8}
\end{array} \right)
\eeq where $M$ is the mass matrix (with no 0-entries) for the
previously mentioned sector 1, while D is a diagonal matrix which is
for the previously mentioned sector 2. So, the sectors are clearly
separated. The sector 1 part, has 3 very complicated eigenvalues,
and zero is eigenvalue with multiplicity of 4. The sector 2 part has
eigenvalues (entries) ordered after the $\xi$-fields:
$(0,0,2A^2g_3^2,2A^2g_3^2,2A^2g_3^2,2A^2g_3^2,2(1+A^2)g_2^2,2(1+A^2)g_2^2)\varphi^2$.
Also, it is important to notice that M is time-dependant. All the
elements involving 6 or 7 are time dependent.

The elements (symmetry implied) are \bea
M_{1,1}&=&4g_1^2,M_{1,2}=-2g_1^2,M_{1,3}=-2\sqrt{1+A^2}g_1^2,M_{1,4}=\frac{2Ag_1^2}{3},M_{1,5}=\frac{4Ag_1^2}{3}\nonumber\\
\nonumber M_{2,2}&=&g_1^2+g_2^2,M_{2,3}=\sqrt{1+A^2}(g_1^2-g_2^2),M_{2,4}=A((-g_1^2/3)+g_2^2),M_{2,5}=\frac{-2Ag_1^2}{3}\\
\nonumber M_{3,3}&=&(1+A^2)(g_1^2+g_2^2),M_{3,4}=-\frac{1}{3}A\sqrt{1+A^2}(g_1^2+3g_2^2),M_{3,5}=-\frac{2}{3}A\sqrt{1+A^2}g_1^2\\
\nonumber M_{4,4}&=&\frac{A^2(g_1^2+9g_2^2+12g_3^2)}{9},M_{4,5}=\frac{2(g_1^2-6g_3^2)A^2}{9},M_{5,5}=\frac{4A^2(g_1^2+3g_3^2)}{9}\\
\nonumber M_{1,6}&=&\frac{2g_1^2\left((1+A^2)\sin[\frac{(8(-1+A^2)\ga)}{5(1+A^2)}-2\si]+\sqrt{1+A^2}\left(-A\sin[\frac{4((3+2A^2)\ga)}{5(1+A^2)}-2\si]+\sin[{\frac{4(2+3A^2)\ga}{5(1+A^2)}+2\si}]\right)\right)}{\sqrt{5}\sqrt{1+A^2}}\\
\nonumber
M_{1,7}&=&\frac{\sqrt{\frac{2}{5}}g_1^2\left(2(1-A^4)\sin[\frac{8(-1+A^2)\ga}{5(1+A^2)}-2\si]+\sqrt{1+A^2}\left(A(3+2A^2)\sin[\frac{4(3+2A^2)\ga}{5(1+A^2)}-2\si]+(2+3A^2)
\sin[\frac{4(2+3A^2)\ga}{5(1+A^2)}+2\si]\right)\right)}{\sqrt{3+7A^2+7A^4+3A^6}}\\
\nonumber
M_{2,6}&=&-\frac{(g_1^2-g_2^2)\left((1+A^2)\sin[\frac{8(-1+A^2)\ga}{5(1+A^2)}-2\si]+\sqrt{1+A^2}\left(-A\sin[\frac{4(3+
2A^2)\ga}{5(1+A^2)}-2\si]+\sin[\frac{4(2+3A^2)\ga}{5(1+A^2)}+2\si]\right)\right)}{\sqrt{5}\sqrt{1+A^2}}\\
M_{2,7}&=&\frac{(g_1^2-g_2^2)\left(2(A^4-1)\sin[\frac{8(-1+A^2)\ga}{5(1+A^2)}-2\si]-\sqrt{1+A^2}\left(A(3+
2A^2)\sin[\frac{4(3+2A^2)\ga}{5(1+A^2)}-2\si]+(2+3A^2)\sin[\frac{4(2+3A^2)\ga}{5(1+A^2)}+2\si]\right)\right)}{\sqrt{10}\sqrt{3+7A^2+7A^4+3A^6}}\\
\nonumber
M_{3,6}&=&-\frac{(g_1^2+g_2^2)\left((1+A^2)\sin[\frac{8(-1+A^2)\ga}{5(1+A^2)}-
2\si]+\sqrt{1+A^2}((-A\sin[\frac{4((3+2A^2))\ga}{5(1+A^2)}-2\si]+\sin[\frac{4(2+3A^2)\ga}{5(1+A^2)}+2\si]))\right)}{\sqrt{5}}\\
\nonumber
M_{3,7}&=&-\frac{(g_1^2+g_2^2)\left(2(1-A^4)\sin[\frac{8(-1+A^2)\ga}{5(1+A^2)}-2\si]+\sqrt{1+A^2}\left(A(3+2A^2)\sin[\frac{4(3+2A^2)\ga}{5(1+A^2)}-2\si]
+(2+3A^2)\sin[\frac{4((2+3A^2))\ga}{5(1+A^2)}+2\si)\right)\right)}{\sqrt{10}\sqrt{3+4A^2+3A^4}}\\
\nonumber
M_{4,6}&=&\frac{A(g_1^2+3g_2^2)\left((1+A^2)\sin[\frac{8(-1+A^2)\ga}{5(1+A^2)}-2\si]+\sqrt{1+A^2}\left(-A\sin[\frac{4(3+2A^2)
\ga}{5(1+A^2)}-2\si]+\sin[\frac{4(2+3A^2)\ga}{5(1+A^2)}+2\si]\right)\right)}{3\sqrt{5}\sqrt{1+A^2}}\\
\nonumber
M_{4,7}&=&\frac{A(g_1^2+3g_2^2)\left(2(1-A^4)\sin[\frac{8(-1)+A^2)\ga}{5(1+A^2)}-2\si]+
\sqrt{1+A^2}\left(A(3+2A^2)\sin[\frac{4(3+2A^2)\ga}{5(1+A^2)}-2\si]+(2+3A^2)\sin[\frac{4(2+3A^2)\ga}{5(1+A^2)}+2\si]\right)\right)}
{3\sqrt{10}\sqrt{3+7A^2+7A^4+3A^6}}\\
\nonumber
M_{5,6}&=&\frac{2Ag_1^2\left((1+A^2)\sin[\frac{8(-1+A^2)\ga}{5(1+A^2)}-2\si]+\sqrt{1+A^2}\left(-A\sin[\frac{4(3+2A^2)\ga}{5(1+
A^2)}-2\si]+\sin[\frac{4(2+3A^2)\ga}{5((1+A^2)}+2\si]\right)\right)}{3\sqrt{5}\sqrt{1+A^2}}\\
\nonumber
M_{5,7}&=&\frac{\sqrt{\frac{2}{5}}Ag_1^2\left(-2(-1+A^4)\sin[\frac{8(-1+A^2)\ga}{5(1+A^2)}-2\si]
+\sqrt{1+A^2}\left(A(3+2A^2)\sin[\frac{4(3+2A^2)\ga}{5(1+A^2)}-2\si]+(2+3A^2)\sin[\frac{4(2+3A^2)\ga}{5(1+A^2)}+2\si]\right)\right)}{3\sqrt{3+7A^2+7A^4+3A^6}}\eea

$M_{6,6},M_{6,7},M_{7,7}$ are much more complicated and are omitted
here.

\section{$QLD^c+LLE^c$ - preheating}\label{sQLDLLE}
In this case, where the mass matrix is time dependent, we must redo
eq.$\ref{Gammaeq}$ and we find \bea
C^T\dot{C}=B^TA^T(\dot{A}B+A\dot{B})=-B^TUB+B^T\dot{B} \eea and
therefore \bea\label{jmatrixQLDLLE^c}
J&=&\frac{1}{2}\left(\sqrt{\omega}\,C^T\dot{C}\frac{1}{\sqrt{\omega}}-\frac{1}{\sqrt{\omega}}\,C^T\dot{C}\sqrt{\omega}\right)\nonumber\\
&=&\frac{1}{2}\left(\sqrt{\omega}\,(-B^TUB)\frac{1}{\sqrt{\omega}}-\frac{1}{\sqrt{\omega}}\,(-B^TUB)\sqrt{\omega}\right)
+\frac{1}{2}\left(\sqrt{\omega}\,B^T\dot{B}\frac{1}{\sqrt{\omega}}-\frac{1}{\sqrt{\omega}}\,B^T\dot{B}\sqrt{\omega}\right)\\
 &=&J_1+J_2 (\textbf{with the obvious definition}).\nonumber\eea

It has been shown numerically that $J_1$ has the following structure
 - treating everything but $\xi_i,\si,\ga$ as constants -
\bea J_1= \left(\begin{array}{ccccccccccccccc}
0&0&0&NZ&NZ&NZ&NZ&0&0&0&0&0&0&0&0\\
0&0&0&NZ&NZ&NZ&NZ&0&0&0&0&0&0&0&0\\
0&0&0&NZ&NZ&NZ&NZ&0&0&0&0&0&0&0&0\\
NZ&NZ&NZ&0&0&0&0&0&0&0&0&0&0&0&0\\
NZ&NZ&NZ&0&0&0&0&0&0&0&0&0&0&0&0\\
NZ&NZ&NZ&0&0&0&0&0&0&0&0&0&0&0&0\\
NZ&NZ&NZ&0&0&0&0&0&0&0&0&0&0&0&0\\
0&0&0&0&0&0&0&0&0&0&0&0&0&0&NZ\\
0&0&0&0&0&0&0&0&0&0&0&0&0&NZ&0\\
0&0&0&0&0&0&0&0&0&0&0&0&0&0&0\\
0&0&0&0&0&0&0&0&0&0&0&0&0&0&0\\
0&0&0&0&0&0&0&0&0&0&0&0&0&0&0\\
0&0&0&0&0&0&0&0&0&0&0&0&0&0&0\\
0&0&0&0&0&0&0&0&NZ&0&0&0&0&0&0\\
0&0&0&0&0&0&0&NZ&0&0&0&0&0&0&0
\end{array} \right)
\eea where NZ stands for a nonzero element. Since the columns of $B$
were chosen such that the first 3 columns represent the massive
states of sector 1, the next 4 the massless states in sector 1, and
the last 8 columns represent the states of sector 2 - in the order
in which the eigenvalues were mentioned, this represents particle
production from the rotation between the 3 massive states and the
massless states of sector 1, and particle production from the
rotation between 2 of the massive and the massless states of sector
2. Reassuringly, there is no particle producing rotations between
(indistinguishable) states of the same mass. Also it is easy to see
that $J_2$ will not alter this picture, since the last 8 columns of
$B$ are constant. This make the last 8 columns of $\dot{B}$ (and
therefore of $B^T\dot{B}$) zero and since multiplying by diagonal
matrices cannot chance zero entries, it is clear that at least the
last 8 columns of $J$ are identical to the last 8 columns of $J_1$.
Therefore the J-matrix is definitely nonzero, and there is particle
production from the rotating eigenstates. It was shown in \cite{us}
that in general all phases will have nontrivial dynamics, which is
necessary for the conclusion that $J$ is nonzero.

\section{the method of field counting - and its limits}\label{scfa}
A simpler approach for determining if preheating is possible is to
count the fields, establishing the number of broken generators and
thus the number of Goldstones and the number of Higgses and by
subtraction finding the number of remaining, physical light degrees
of freedom \cite{Olive:2006uw}. This can be done sector by sector.

$LLE^c$ breaks $SU(2)\times U(1)$ completely. It has 6 fields (real)
in sector 1, it breaks 2 diagonal generators and thus have 2 Higgses
and 2 Goldstones. This leaves 2 light degrees of freedom, which
corresponds to the flat direction. (It is clear that the sum of the
phases cannot be gauged away.) However, one needs to argue why the 2
Higgses cannot rotate between each other - since they have different
eigenvalues a possible rotation would be physical, and why the flat
direction stays out of this rotation. In this case the flat
direction - understood as 2 real fields, that is the direction
itself, and the field combination that is orthogonal to it in all
superfields individually - stays constant. So it is clearly not
rotating. Sector 2 is very easy. There are 4 fields, and 2 broken
off-diagonal generators and therefore 2 Higgses and 2 Goldstones.
Since the 2 Higgses have the same eigenvalue, there will surely not
be particle production in this sector.

$U^cD^cD^c$ breaks $SU(3)_c\times U(1)$ to $U(1)_{NEW}$. It has 6
fields in sector 1, it breaks 2 diagonal generators and thus have 2
Higgses and 2 Goldstones. This leaves 2 light degrees of freedom,
which corresponds to the flat direction.  However, as before, one
needs to argue why the 2 Higgses cannot rotate between each other.
The flat directions stay out of this rotation. Sector 2 is again
very easy. There are 12 fields, and 6 broken off-diagonal generators
and therefore 6 Higgses and 6 Goldstones. Since the 2 Higgses have
the same eigenvalue, there will surely not be particle production in
this sector.

$QLQLQLE^c$ breaks $SU(3)_c\times SU(2)\times U(1)$ completely. It
has 14 fields in sector 1, it breaks 4 diagonal generators and thus
have 4 Higgses and 4 Goldstones. This leaves 6 light degrees of
freedom, corresponding to the 1 flat direction and 4 additional
light degrees of freedom to which there can be rotations which give
preheating. In Sector 2 there are 38 fields, and 8 broken
off-diagonal generators and therefore 8 Higgses and 8 Goldstones.
Since each field is exclusively connected to the VEV by $SU(3)$ or
$SU(2)$ the Higgses cannot rotate between each other. 12 fields are
completely decoupled (those of Q, differing in both $SU(3)_c$ and
 $SU(2)$-charge from the VEV).Indeed there is rotation to some of
 the remaining 10 states, but it is hard to argue exactly why and to
 how many, without doing the full investigation.

$LLE^c+U^cD^cD^c$ breaks $SU(3)_c\times SU(2)\times U(1)$
completely. It has 12 fields in sector 1, it breaks 4 diagonal
generators and thus have 4 Higgses and 4 Goldstones. This leaves 4
light degrees of freedom, corresponding to the 2 flat directions.
Again one needs to argue why the 4 Higgses cannot rotate between
each other. The flat direction clearly stays out of the rotation. In
Sector 2 there are 16 fields, and 8 broken off-diagonal generators
and therefore 8 Higgses and 8 Goldstones. Since each field is
exclusively connected to the VEV by $SU(3)$ or $SU(2)$ the Higgses
cannot rotate between each other.

$QLD^c+LLE^c$ breaks $SU(3)_c\times SU(2)\times U(1)$ to $SU(2)_c$.
It has 10 fields  in sector 1 (one complex field in common), it
breaks 3 diagonal generators and thus have 3 Higgses and 3
Goldstones. This leaves 4 light degrees of freedom, corresponding to
the 2 flat directions. Again one needs to argue why the 3 Higgses
cannot rotate between each other and why in this case the light
fields corresponding to the flat directions \emph{does rotate} with
the Higgses. In Sector 2 there are 18 fields and 6 broken
off-diagonal generators and therefore 6 Higgses and 6 Goldstones and
4 are completely decoupled. This leaves 2 light fields that can
rotate. One can also argue, that the 4 down fields in $Q$ and 4
strange fields in $D^c$ must represent the 4 color Higgses and 4
color Goldstones with no particle production.

It seems clear, that while this counting is a nice tool to look for
opportunities for preheating and to exclude preheating especially in
sector 2, it is still necessary to do the full analysis in the
unitary gauge to draw firm conclusions - at least when it comes to
the role of the fields corresponding to the flat directions
themselves.

\section{Summary and conclusion}\label{scon}
For the conclusions on one flat direction, see section \ref{sOFDs}.
For 2 flat directions, we have found that particle production is
possible in $QLD^c+LLE^c$ but not in $U^cD^cD^c+LLE^c$. The
difference seems to be the presence of a common field in the former
case, but not in the latter. We have also found that it is necessary
to transform to the unitary gauge after identifying the Goldstones,
in order to make correct conclusions. Counting fields and broken
generators can give hints to whether there is particle production or
not, but it is not sufficient for firm conclusions.

Finally, we shall stress that what we have shown is that there will
be particle production in the $QLD^c+LLE^c$ case. However, the
statement that both directions are likely to get large VEV's
\cite{Olive:2006uw} has not been investigated in this paper. Neither
has the very recent claim that even if non-perturbative particle
production happen, the main decay mode will still be perturbative
\cite{Allahverdi:2008}. Also, whether the rotation of the flat
directions are fast enough for this particle production to lead to
preheating and thus not giving the effect of delayed thermalisation
is outside the scope of the present paper. We presume that the
situation is close to the situation in \cite{us} - and that there
will be very significant particle production. However, as stated in
\cite{us}, the effect of SUSY breaking terms in the Lagrangian has
not been taken into account.

But we can conclude that in order to determine the role of SUSY flat
directions in (p)reheating, it is absolutely necessary to determine
which flat directions get the large VEVs (only a limited number of
the countless flat directions can get large VEVs at the same time)
and if many directions get a large VEV a numerical study will
probably be necessary to determine if the role of some additional
flat directions can be ignored.

\section{Acknowledgements}
I would like to thank David Maybury, Francesco Riva and Stephen M
West for the collaboration that led to the formalism used in the
present paper. I would also like to thank Steen Hannestad and Martin
S Sloth for useful discussions.

\end{document}